\documentstyle[twoside,fleqn,espcrc2,epsf]{article}

\newcommand{\AmS}{{\protect\the\textfont2
  A\kern-.1667em\lower.5ex\hbox{M}\kern-.125emS}}

\title{Exclusive production of dijets in QCD}

\author{D.Yu.~Ivanov
\address{Institut 
f\"ur Theoretische Physik, Universit\"at
          Regensburg, D-93040 Regensburg, Germany }
\address{Institute of Mathematics, 630090 Novosibirsk, Russia}
        \thanks{
This work was supported by Alexander von Humboldt Foundation.}}

\begin{document}

\begin{abstract}
We study in a framework of QCD collinear factorization 
the processes of a pion and a photon
diffraction dissociation into two jets.
The structure of non-factorizable contributions 
is discussed. 
We argue that coherent production of hard dijets by linearly
polarized real photons can provide direct
evidence for chirality violation in hard processes, the
first measurement of the magnetic susceptibility of the quark condensate
and the photon distribution amplitude.
\end{abstract}

\maketitle

\section{Introduction}
The pion (and photon) diffraction dissociation into a
pair of jets with large transverse momentum on a nucleon target
was first discussed in \cite{KDR80}. The possibility
to use this process to probe the nuclear filtering of pion components
with a small transverse size was suggested in \cite{BBGG81}.
The A-dependence and the
$q_\perp^2$-dependence of the
coherent dijet cross section was first calculated
in \cite{FMS93} and it was argued that the jet
distribution with respect to  the longitudinal momentum fraction has to
follow the quark momentum distribution in the pion and hence provides
a direct measurement
of the pion distribution amplitude. Recent experimental data
by the E791 collaboration \cite{E791a} indeed confirm the strong
A-dependence which is  a signature for color transparency, and are
consistent
with the predicted $\sim 1/q_\perp^8$
dependence on the jet transverse momentum. Moreover, the jet longitudinal
momentum fraction distribution turns out to be consistent with the
$\sim z^2(1-z)^2$ shape corresponding to the asymptotic pion distribution
amplitude. 
After these first successes, one naturally asks whether the
QCD description of coherent dijet production can be made fully quantitative.

We developed a perturbative QCD framework for the description
of coherent dijet production that is in line with other known
applications of the QCD factorization techniques.
First we discuss factorization and 
concentrate on a pion dissociation process, then  
we will consider production of dijets initiated by a real photon,
a process which  
is sensitive to the chiral--odd properties of QCD vacuum.
The results reported here have been obtained in collaboration with
V. Braun, S. Gottwald, A. Sch\"afer and L. Szymanowski 
\cite{BISS01,Braun:2002en}.

\section{Pion dissociation}

\begin{figure}[t]
\epsfxsize7.5cm\epsffile{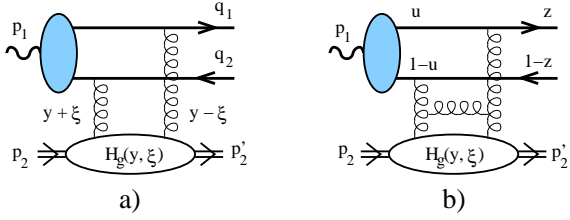}
\vskip-1.0cm
\caption{Sample diagrams for the hard dijet production, see text.
}
\label{fig:1}
\end{figure}

The kinematics of the process and the notation for the
momenta is shown in Fig.~\ref{fig:1}.
The Sudakov decomposition of the jet momenta with respect to
the momenta of the incoming particles $p_1$ and $p_2$ reads
\begin{equation}
q_1=zp_1+\frac{q_{\perp}^2}{zs}p_2+q_{\perp}\,,
q_2=\bar zp_1+\frac{q_{\perp}^2}{\bar zs}p_2-q_{\perp}
\label{Sudakov}
\end{equation}
so that $z$ is the longitudinal momentum fraction
and $q_\perp$ the transverse momentum of the quark
jet.
Hereafter
we use the shorthand notation $\bar u \equiv (1-u)$ for any
longitudinal momentum fraction $u$.
Note that we consider the forward limit,
when transverse momenta of the jets compensate each other.
In this kinematics the invariant mass of the produced $q\bar q$ pair is
equal to
$
M^2={ q_{\perp}^2}/{ z\bar z}
$,
and the momentum of the outgoing nucleon $p_2^\prime=p_2(1-\xi)/(1+\xi)$,
where
$\xi= M^2/(2s-M^2) \simeq M^2/2s$, $s=(p_1+p_2)^2$.
%

The possibility to constrain the pion distribution amplitude
$\phi_\pi (u)$ in the dijet diffractive dissociation
experiment assumes that 
the amplitude
\newpage 

\vspace*{-1.0cm}

\noindent 
of this process can be calculated in the collinear
approximation as
suggested by  Fig.~\ref{fig:1}:   
\begin{equation}
{\it M}
=\int\limits^1_0 du \int\limits^1_{-1} dy
\,\phi_\pi
(u)
\,T^g_H (u , y)\,{\cal H}_g(y,\xi )\,.
\label{factor}
\end{equation}
Here ${\cal H}_g(y,\xi)$ is the generalized gluon
distribution in the target nucleon
\cite{Ji97a};
variable $-1<y<1$ parametrizes the
momentum fractions of the emitted and the absorbed gluons.
$T_H^g(u, y)$ is the hard scattering amplitude involving at
least one hard gluon exchange. 

There are two important regions in the integral (\ref{factor}), see 
\cite{BISS01} for more details. At 
$u\to 0,1$ 
\begin{equation}         
 {\it  M}\Big|_{\rm end-points}
  \sim 
i  z \bar z \,  \int\limits^1_{u_{\rm  min}} du\,
\frac{\phi_\pi(u)}{u^2}{\cal H}_g(\xi,\xi) \,.
\label{end}
\end{equation}
Since $\phi_\pi(u)\sim u$ at $u\to 0$, the integral over $u$
diverges logarithmically.
Remarkably, the integral containing the pion distribution
amplitude does not involve any $z$-dependence.
Therefore, the longitudinal momentum distribution of the jets in the
nonfactorizable contribution is calculable and, as it turns out,
has the shape of the asymptotic pion distribution amplitude
$\phi_\pi^{\rm as}(z) = 6z\bar z$.

In technical terms, the appearance of the end point divergence
is due to pinching of the $y$ contour at the point $y=\xi$
in case that the variable $u$
is close to the end--points.
One can trace \cite{BISS01} that this pinching occurs
between soft gluon 
interactions in the initial and in the final state, and is related with
the existence of the unitarity cuts of the amplitude in different,
$s$ and $M^2$, channels.

The other important integration region in (\ref{factor}) is the one 
when the longitudinal momentum fraction carried by the quark is close
(for high energies) to that of the quark jet in the final state
\begin{equation}       
{\it M}\Big|_{\xi\ll |u- z|\ll 1}
\sim 4 i \phi_\pi(z)\!\int\limits_\xi^1
 \!\frac{dy}{y+\xi}\, {\cal H}_g(y,\xi)\,.
\label{z=z'}
\end{equation}
This logarithmic integral is nothing but
the usual energy logarithm that accompanies each extra gluon in the
gluon ladder. Its appearance is due to the fact the hard gluon
which supplies jets by the high transverse momentum
can be emitted in a broad rapidity interval and
is not constrained to the pion fragmentation region.
The integral on the r.h.s. of (\ref{z=z'}) can be identified with the
unintegrated generalized gluon distribution.
Therefore, in this region hard
gluon exchange can be viewed as a large transverse momentum part of the
gluon distribution in the proton, cf. \cite{NSS99}.
This contribution is proportional to
the pion distribution amplitude $\phi_\pi (z)$
and contains the enhancement factor $\ln 1/\xi \sim \ln s/q_\perp^2$.

\section{Photon dissociation}

The wave function of a real photon contains both the
perturbative chiral-even (CE) contribution of
the quark-antiquark pair with opposite helicities, and the nonperturbative
chiral-odd (CO) contribution with quarks having the same helicity and which
is due to the chiral symmetry breaking. It is proportional to fundamental
parameters of QCD vacuum, quark condensate $\langle \bar q q \rangle$ 
and magnetic susceptibility $\chi$.  
The perturbative CE contribution is
singular $\sim 1/|{\bf r}|$ at small transverse distances ${\bf r}$. 
The nonperturbative CO contribution is regular at small transverse
separations
and can be parametrized by the photon
distribution
amplitude $\phi_\gamma(u,\mu)$ \cite{BBK89}
\begin{eqnarray}
\lefteqn{\langle 0 |\bar q(0) \sigma_{\alpha\beta} q(x)
   | \gamma^{(\lambda)}(q)\rangle = i \,e_q\, \chi\, \langle \bar q q \rangle
\times }
\label{phigamma}\\
&& \left( e^{(\lambda)}_\alpha q_\beta-  e^{(\lambda)}_\beta q_\alpha\right)
 \int\limits_0^1 \!du\, e^{-iu(qx)}\, \phi_\gamma(u,\mu)\,.
 \nonumber
\end{eqnarray}
$\phi_\gamma(u, \mu \ge 1~\mbox{\rm GeV})$
is believed to be not far from the asymptotic form $
   \phi_\gamma^{\rm as}(u) = 6 u (1-u)$.
$\chi$ was estimated using the vector dominance
approximation and QCD sum rules~\cite{chiold,chi}:
$\chi \langle \bar q q \rangle \simeq 40-70~\mbox{\rm  MeV}$.
However, any direct experimental evidence on both $\chi$ and
$\phi_\gamma(u)$
is absent. 

 
This structure can be studied in experiments
similar to the studies of coherent dijets in  
pion dissociation by the E791 collaboration \cite{E791a}. 
The approach of \cite{E791a} is different as compared to
to earlier studies of the dijet photoproduction \cite{H1}
in that the exclusive dijet final state is
identified by requiring that the jet transverse momenta are compensated
to a high accuracy within the diffractive cone and making some additional
cuts. This approach seems to work for the case of pion dissociation, 
and for photoproduction the corresponding experimental program  
is under way at HERA \cite{Ashery02}. In that case one should take care about 
the production of heavy quark dijets since for heavy quarks there is 
a perturbative chiral--odd effect due to a quark mass. 
We assume that separation of the exclusive light $q\bar q$
dijet final state is feasible.

Since the CE and CO contributions lead to final states with
different helicity, they do not interfere and the dijet
cross section is given by the incoherent sum, for the linearly polarized
photon
\begin{eqnarray}
\lefteqn{ \frac{d\sigma_{\gamma\to 2\,{\rm jets}}}{d\phi
d q_\perp^2 dt dz}{\Bigg|}_{t=0}
\ =\ \sum_q e^2_q \frac{\alpha_{EM} \alpha_s^2 \pi^2
(1+\xi )^2
}{4\pi N_c q_\perp^6}\times } \\
\lefteqn{\left[(1-4z\bar z \cos^2\!{\phi})
|{\cal J}_{CE}|^2+\frac{\pi^2 \alpha_s^2 \chi^2 \langle
\bar q q \rangle^2 }{N_c^2q^2_\perp}|{\cal J}_{CO}|^2\right], }\nonumber
\label{c.sect.}
\end{eqnarray}
where $\phi$ is the azimuthal angle between the jet direction and
the photon polarization $(e^{(\lambda)}\cdot q_\perp) \sim \cos\phi$,
${\cal J}_{CE}$ and ${\cal J}_{CO}$ are the CE and CO amplitudes
respectively.
Note that the CE contribution is $\sim 1/q_\perp^6$ \cite{FMS93} and the
CO contribution is suppressed by one extra
power of $q_\perp^2$ which follows from twist counting.
The different $\phi$ dependence can be traced to the fact that
the $q\bar q$ pair is produced in a state with orbital angular momentum
$L_z=0$ and $L_z=\pm 1$ for the CO and CE contributions, respectively.

The CE contribution originates from the region of large momenta flowing
through
the photon vertex and was considered previously in
\cite{NNN,BLW96,LMRT97,GKM98}
in the high energy limit using $k_\perp$ factorization.
We believe that the collinear factorization is more adequate for HERA
energies an
in difference to the dijet production by incident pions, expect
that
it is valid for the CE amplitude to all orders in perturbation theory.
To leading order (LO) in the strong coupling $\alpha_s = \alpha_s(q_\perp)$
the amplitude is given by the sum of Feynman diagrams of the type
shown in Fig.~\ref{fig:1}a with all possible attachments of the gluons.
The answer reads
\begin{equation}
{\cal J}_{CE}= i \xi {\cal H}_{g}^\prime (\xi,\xi)
+\frac{i \alpha_s N_c}{\pi}\int^1_\xi\frac{dy }{y+\xi}
{\cal H}_{g} (y,\xi)\,,
\label{even}
\end{equation}
where ${\cal H}_{g}^\prime (\xi,\xi)=d{\cal H}_{g} (y,\xi)/dy|_{y=\xi}$.
The second term in eq.~(\ref{even}) originates from the diagrams with
additional gluon exchange between the
$t$-channel gluons, see Fig.~\ref{fig:1}b, it corresponds to the leading 
at
large energies  
(enhanced by
$\log\xi$) NLO
contribution.
At $y\to 0$, \,
${\cal H}_{g}(y,\xi)\sim y^{-\Delta}$, where  in perturbation theory $\Delta
\sim
\alpha_s\log{s/q^2_\perp}$ has to be treated as
a small parameter. Therefore, despite the fact that the two terms in
(\ref{even}) appear in different orders in the collinear expansion,
they are of the same order as far as the counting of energy logarithms
is concerned. 
Since ${\cal H}_{g} (y,\xi)\sim
G(y)$ at $y\gg \xi$, and as the factor
$\alpha_s N_c/(\pi y)$ is nothing but the low-$y$ limit of the
DGLAP gluon splitting function,  the integral in Eq.~(\ref{even}) can be
identified to logarithmic accuracy with the unintegrated gluon distribution
$f(\xi,q^2)=\partial G(\xi,q^2)/\partial \ln q^2$. This contribution
corresponds
one considered in \cite{NNN,BLW96,LMRT97} in the $k_\perp$ factorization
approach
The first contribution in (\ref{even}) is analogous to Eq.~(42) in
\cite{GKM98}.

For the nonperturbative CO contribution
the large momenta are not allowed in the
photon vertex and the factorization formula contains a
convolution with the photon distribution amplitude.
In this case an additional hard gluon exchange is mandatory and the
diagram in Fig.~\ref{fig:1}b presents one example of the existing 31 LO
contributions. Similar to the pion case 
we found that the result for the amplitude may be  
approximated well by the sum of two contributions in analogy to 
eqs.~(\ref{end}) and (\ref{z=z'}).   
The origin of the end--point divergence 
is the same as in a pion dissociation.
In the present context violation of factorization is probably
not surprising since the CO contribution is suppressed by a power of
$q_\perp^2$
compared to the leading twist.
Assuming that the photon distribution amplitude
is close to the asymptotic form,
we obtain ${\cal J}_{CO}\sim z(1-z)$ for both integration regions, up to
small
corrections. The presence of 
nonfactorizable contribution does not have, therefore,
any significant effect on the jet distribution but mainly influences the
normalization.


In the numerical calculation performed for  HERA kinematics we 
have introduced an infrared cutoff 
$u_{\rm  min} = \mu^2_{\rm
IR}/q_\perp^2$ to regularize the nonfactorizable contribution (\ref{end}),
$\mu_{\rm IR}=500$~MeV.
We found that the nonperturbative CO contribution
integrated over $\phi$, $z$ and
$t$
is of the order
of
\begin{equation}
 \frac{d\sigma_{CO}}{d\sigma_{CE}} \simeq
   (7\pm 2~\mbox{\rm GeV})^2\cdot \frac{\alpha_s(q_\perp)^2}{q_\perp^2}\,
   \left(\frac{\chi\langle \bar q q\rangle}{50~\mbox{\small \rm
MeV}}\right)^2.
\label{COscale}
\end{equation}
For $q_\perp > 4$~GeV the cross section is dominated by the perturbative CE
contribution, for smaller transverse momenta the
dijet cross
section is saturated by the CO contribution.
The transition between the two different regimes 
is seen very clearly from
the dependence of the cross section on the dijet longitudinal momentum
fraction a
the azimuthal angle.
At $q_\perp >4 $~GeV the $z$-distribution is almost flat,
while
the $\phi$ distribution is almost purely $\sim 1-\cos^2\phi$. In contrast to
this
at $q_\perp < 4$~GeV the $z$-distribution is comparable with  $\sim
z^2(1-z)^2$
while the $\phi$-distribution becomes flat.

One can conclude that
studies of exclusive photoproduction
of light quark dijets with large transverse momenta can yield important
information on the photon structure at small distances.
Our main result is that the nonperturbative CO contribution is large in the
region
of intermediate $q_\perp\sim 2-4$~GeV and can be clearly separated from the
perturbative contribution by a different $z$- and $\phi$-dependence.
Observation of the CO contribution would be the first clear
evidence for the chirality violation in hard processes and also provide the
first direct measurement of the magnetic susceptibility of the quark
condensate.
On the other hand, the dijet cross section for large $q_\perp$ can serve to
constrain the generalized gluon distribution.

On the theoretical side, we deviate from previous studies of the dijet
production by consistently applying the collinear factorization
in terms of generalized parton distributions. For the nonperturbative CO
contribution
the collinear factorization is, strictly speaking, broken. However, the
sensitivity
to the infrared cutoff is relatively weak and can formally be eliminated by taking
into
account Sudakov-type corrections in the modified collinear factorization
framework.

\end{document}